\documentclass[prd, twocolumn, nofootinbib, floatfix]{revtex4}

\usepackage{epsfig}
\usepackage{amsmath}

\newcommand{\beq}{\begin{equation}}
\newcommand{\eeq}{\end{equation}}
\newcommand{\beqa}{\begin{eqnarray}}
\newcommand{\eeqa}{\end{eqnarray}}

\newcommand{\lam}{\lambda}
\newcommand{\gam}{\gamma}

\newcommand{\ls}{\mathrel{\raise0.27ex\hbox{$<$}\kern-0.70em \lower0.71ex\hbox{{
$\scriptstyle \sim$}}}}

\begin{document}

\title{Cosmological Constant Behavior in DBI Theory}
\author{Changrim Ahn${}^{1}$, Chanju Kim${}^{1}$, and Eric V.\ Linder${}^{1,2}$}
\affiliation{${}^1$ Institute for the Early Universe and Department 
of Physics, Ewha Womans University,
Seoul 120-750, South Korea\\
${}^2$ Berkeley Center for Cosmological Physics and Berkeley Lab, 
University of California, Berkeley 94720, USA}
\date{\today}

\begin{abstract}
Cosmological constant behavior can be realized as solutions of the
Dirac-Born-Infeld (DBI) action within Type IIB string theory and the
AdS/CFT correspondence.  We derive a family of attractor solutions
to the cosmological constant that arise purely from the ``relativistic''
nature of the DBI action without an explicit false vacuum energy.
We also find attractor solutions with values of the equation
of state near but with $w\ne-1$; the forms for the potential arising
from flux interactions are renormalizable and natural, and the
D3-brane tension can be given by the standard throat form.
We discuss present and future observational constraints on the theory. 
\end{abstract}

\maketitle

\section{Introduction \label{sec:intro}}

The cosmological constant is a fundamental puzzle for high energy
physics field theory.  With the observational discovery of the
accelerated expansion of the universe \cite{perl99,riess98}, this
puzzle has become a premier challenge for both theoretical and
observational physics.  The cosmological constant, or something with
similarly negative effective pressure, dominates the energy density
of the universe.  However, it is not at all clear how it arises
naturally within a fundamental theory of physics.  In particular,
the needed energy density lies $10^{121}$ times below the Planck
energy density, requiring severe fine tuning of high energy physics.

Giving dynamics to the field allows the possibility
of ameliorating the fine tuning issue, through scalar fields that
exhibit attractor behavior.  However, it has been quite difficult to
devise fields that both have attractors and can achieve sufficiently
negatively equations of state (pressure to energy density ratios) by
the present epoch.

String theory can impose a non-trivial kinetic
behavior through the Dirac-Born-Infeld (DBI) action that arises naturally
in consideration of D3-brane motion within a warped compactification.
Several articles have considered DBI as a source for inflation
\cite{silvertong,ast,chimentolazkoz} or dark energy
\cite{martinyamaguchi}, fixing
one or another function within the DBI action.
We focus on the unusual dynamics (following the pioneering work of
\cite{silvertong} for inflation) and find this can have several
important consequences with advantages for bringing the theory naturally
into close accord with observations.
In particular, a near cosmological constant state can be achieved
in several ways uniquely distinct from quintessence.

In Section~\ref{sec:eom} we lay out the foundations of DBI theory and
the important parameters in the equations of motion.  We find solutions
insensitive to initial conditions, i.e.\ having a large basin of
attraction, in Section~\ref{sec:attr}, including for cosmological
constant behavior.  The most interesting cases arise uniquely from
the ``relativistic'' nature of the DBI action and have quite simple
potentials.

\section{DBI Methodology \label{sec:eom}}

We consider the low-energy dynamics of a probe D3-brane in a warped geometry
coupled to gravity. It is governed by the DBI action
\cite{silvertong},
\beq
S=-\int d^4x\,\sqrt{-g}\,\left[T(\phi)\sqrt{1-\dot\phi^2/T(\phi)}+V(\phi)-T(\phi)\right]\,,
\eeq
where we ignored the spatial derivatives of $\phi$.
$T$ is the warped brane tension 
and $V$ is the potential arising from interactions with Ramond-Ramond fluxes
or other sectors. 
The energy-momentum tensor
takes a perfect fluid form with energy density $\rho_\phi$ and pressure
$p_\phi$ given by
\beq
\rho_\phi=(\gamma-1)\,T+V \quad ;\quad p_\phi=(1-\gamma^{-1})T-V\,.
\eeq
The Lorentz factor $\gamma$ measures the ``relativistic'' motion
of the field,
\beq
\gamma=(1-\dot\phi^2/T)^{-1/2}\,.
\eeq

The equation of state for the DBI field is
\beq
w\equiv \frac{p_\phi}{\rho_\phi}=-\frac{\gamma^{-1}-1+v}{\gamma-1+v}
\,, \label{eq:wdef}
\eeq
where $v(\phi)=V(\phi)/T(\phi)$.  In the ``nonrelativistic'' limit,
$\gamma\to 1+K/T$, where $K\equiv\dot\phi^2/2$ is the canonical kinetic
energy, and $w\to (K-V)/(K+V)$ as for a quintessence field.  However,
the noncanonical behavior due to the relativistic corrections will be
crucial.

The equation of motion for the field follows from either functional
variation of the action or directly from the continuity equation for
the energy density,
\beq
\rho'_\phi=-3(\rho_\phi+p_\phi)=-3(\gamma-\gamma^{-1})\,T\,,
\eeq
where a prime denotes a derivative with respect to the e-folding parameter,
$d/d\ln a$.  The necessary ingredients are the tension $T(\phi)$ and
potential $V(\phi)$, and initial conditions on the field.  Solving for
the evolution then delivers the equation of state parameters $w(a)$ and
$w'(a)$ for a phase space portrait of the dynamics, and
$\Omega_\phi(a)=(8\pi G\rho_\phi)/(3H^2)$ for the dark energy density
fraction and computation of the Hubble parameter $H(a)=\dot a/a$ and
cosmological distances $d(a)=\int da/(a^2 H)$.

For a pure AdS$_5$ geometry with radius $R$, the warped tension is given by
\beq
T(\phi)=\tau\,\phi^4\,, \label{eq:tphi}
\eeq
with $\tau = 1/(g_s\tilde\lambda)$ where $g_s$ is the string coupling, 
$\alpha'$ is the inverse string tension, 
and $\tilde\lambda=R^4/\alpha'^2$ which is identified as the 't Hooft coupling
in AdS/CFT correspondence.  
In general we do not need to take advantage of further degrees of
freedom by altering the tension function, although other forms for it
lead to similar conclusions as well.  In the next section we find that very
simple, standard forms of the potential, such as $V(\phi)=m^2\phi^2$ or
$V\sim\phi$, have quite interesting behavior.
Thus, there is little arbitrariness or unnaturalness needed to
find results approaching the cosmological constant behavior.

\section{Cosmological Constant and Other Attractors \label{sec:attr}}

Solutions to the equations of motion where no special time is picked
out in the history of the universe have long been of interest as means
to ameliorate the coincidence or fine tuning problems
\cite{ferreirajoyce,copelw,liddlescherrer,zws}.  Attractor solutions avoid
fine tuning in that the dynamical trajectory
of the field lies along a common track despite starting from different
initial conditions.  In general, only highly specific forms of the
potential possess this characteristic in the quintessence case; we
find that this is vastly enlarged in the DBI case and in fact many
standard potentials such as quadratic and quadratic plus quartic
forms exhibit attractor behavior.  We identify the origin of this as
the relativistic limit where the Lorentz boost factor $\gamma$ grows large;
hence it is an innate characteristic of DBI string theory.

To begin, we define the contributions of the tension and potential
to the vacuum energy density relative to the critical density, \beq
x^2=\frac{\kappa^2}{3H^2}(\gamma-1)\,T \quad ; \quad
y^2=\frac{\kappa^2}{3H^2}V \,, \eeq where
$\kappa^2=8\pi G$. 
The equations of motion are given by 
\beqa 
x'&=&-\frac{3}{2\gamma}x(1-x^2)-\frac{3}{2}xy^2+\frac{\sqrt{3}}{2}\lambda\frac{\sqrt{\gamma+1}}{\gamma}y^2\\
y'&=&\frac{3}{2\gamma}x^2y+\frac{3}{2}y(1-y^2)-\frac{\sqrt{3}}{2}\lambda\frac{\sqrt{\gamma+1}}{\gamma}xy
\eeqa 
where $\kappa\phi'=x\sqrt{3(\gamma+1)}/\gamma$, 
$\lam=-(1/\kappa V)dV/d\phi$ and 
\beq 
\gamma=1+\frac{V}{T}\frac{x^2}{y^2}\,. \label{eq:gamv} 
\eeq 

We are interested in the DBI field as late time accelerating dark
energy, not for inflation, so we take the initial conditions in the
matter dominated universe and define the present by
$\Omega_\phi=0.72$. 
The attractor solutions to the equation of motion have the critical values  
\beqa 
x^2_{c1}&=&\frac{\lam^2}{3(\gam+1)} \quad ; \quad 
x^2_{c2}=\frac{3\gamma^2}{\gam+1}\frac{1}{\lam^2} \\ 
y^2_{c1}&=&1-\frac{\lam^2}{3(\gam+1)} \quad ; \quad 
y^2_{c2}=\frac{3\gam}{\gam+1}\frac{1}{\lam^2} \\ 
\Omega_{\phi,c1}&=&1 \quad ; \quad \Omega_{\phi,c2}=\frac{3\gam}{\lam^2} \\ 
w_{\phi,c1}&=&-1+\frac{\lam^2}{3\gam} \quad ; \quad w_{\phi,c2}=0 \,. \label{eq:wc} 
\eeqa 
A key criterion is whether 
$\lam^2/\gam$ is zero, finite, or diverges. For non-negative potentials, 
the first set of critical values only exists for $\lam^2<3\gam$.  The 
second set does not lead to acceleration so we do not consider it  further. 

Note that we have made no assumptions as
to whether $\gam$ or $\lam$ are constant or not in the overall
evolution. To obtain the de Sitter behavior of a cosmological
constant, we need $\lam^2/\gam\to 0$.  This requires either $\lam\to
0$ or $\gam\to\infty$. 

On the attractor $x$, $y$ will be constant so we can write
$\gamma=1+kv$, with $k$ a constant, and a key parameter is $v=V/T$.
One also has that $\gamma'\to(\gam-1)v'/v$ so $\gam$ is driven to
either 1 or $\infty$ (unless $v=\,$constant). Suppose $\gam\to1$.
Then we need $\lam\to0$.  This can be achieved for runaway
potentials (where $\phi\to\infty$) of the inverse power law form,
$V\sim\phi^{-c}$, similar to the quintessence case \cite{ratra}.
For finite values of $\phi$, though, $\lambda\to0$ can only be realized 
for $\phi\to0$ (i.e.\ potentials without poles) by including nonzero 
minimum vacuum energy, i.e.\ an explicit cosmological constant. 
Therefore we turn to the $\gamma\to\infty$ case.

In the fully relativistic, $\gam\to\infty$, limit we can obtain the
cosmological constant behavior.  By Eq.~(\ref{eq:gamv}) this requires
$v\to\infty$.  Suppose we take $T\sim\phi^n$, with $n=4$ giving the
quartic brane tension in AdS space. 
Then following Eq.~(\ref{eq:wc}) 
a simple realization of the
cosmological constant attractor is $V\sim\phi^c$ where $0<c<n-2$.  (Note
that the equations of motion guarantee that the field stops at $\phi=0$
before rolling to negative values of $\phi$, so $c$ is not restricted to
even integers.)

\begin{figure}[!htb]
\begin{center}
\psfig{file=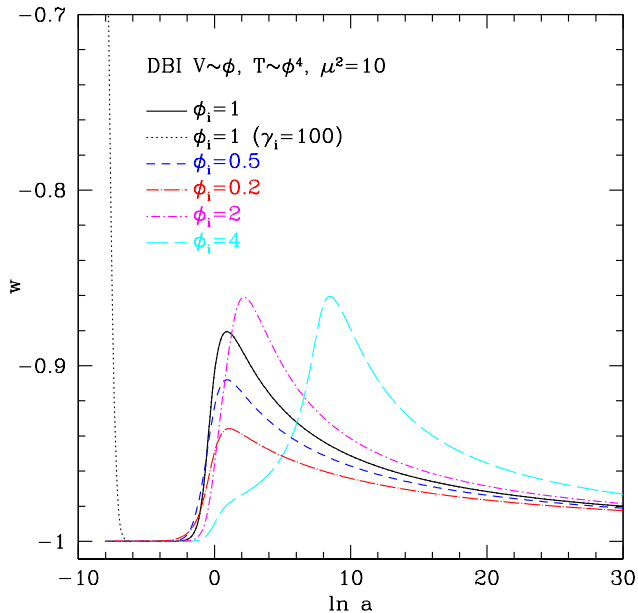,width=3.4in}
\caption{The DBI dynamics can have an attractor to the cosmological
constant state, insensitive to the initial conditions of the field
value $\phi_i$ or boost factor $\gamma_i$.  During the matter dominated
era the field quickly approaches a frozen state with $w=-1$ and
$\gamma=1$, and then thaws as the dark energy density starts to become
appreciable.  In the future, the field joins the attractor solution
with $\gam\to\infty$, and $w\to-1$ as $1/\ln a$.
}
\label{fig:gamlp}
\end{center}
\end{figure}

We illustrate the example of the linear potential, $V\sim\phi$, in
Figure~\ref{fig:gamlp}.  The field indeed goes to the attractor
behavior independent of the initial conditions of the  field value,
$\phi_i$, and field velocity, i.e.\ $\gam_i$ (we discuss the evolution
further in a later section).  At late times the
behavior is just that of a cosmological constant, $w=-1$.

An interesting further point is that we can consider the relativistic
limit but where $\lam^2\to\infty$ also, in such a way that the key
ratio $\lam^2/\gam$ stays finite.  In this case, $w$ approaches an
asymptotic value with $w\ne-1$, but it can lie close to $-1$ and
certainly in the accelerating regime.   This can be realized for
$V\sim\phi^c$ with $c=n-2$.  In particular, a quadratic potential with
quartic tension leads to such a solution.  This is quite interesting
as this is naturally predicted by
DBI theory in pure AdS geometry. 

\begin{figure}[!htb]
\begin{center}
\psfig{file=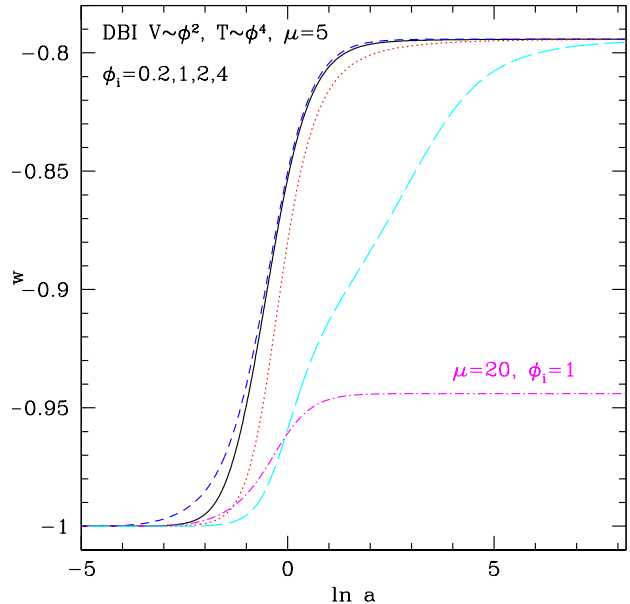,width=3.4in}
\caption{A natural quadratic potential exhibits attractor behavior
in DBI theory due to the relativistic boost factor.  The field
evolves from a frozen $w=-1$ state in the matter dominated era and
goes to a constant equation of state in the future, independent of
initial conditions.  The larger the effective field mass $\mu$, the
closer the behavior approaches a cosmological constant.
}
\label{fig:phi2}
\end{center}
\end{figure}

The potential may arise from the couplings of the D3-brane to fluxes and
other sectors involved in a compactification.
In the case of pure AdS$_5 \times \mathrm{S}^5$ geometry, 
the potential is quartic.  Corrections to the conformal invariance,
however, generically create a mass term giving a quadratic contribution
\cite{ast,silvertong}.  In fact, all we require is that the potential
looks quadratic near its minimum -- a highly generic state.  In the
$c=n-2$ case, the equation of state 
has a negative value 
\beq
w=-1+\frac{c^2}{6\mu^2}\,\left[-1+\sqrt{1+12\mu^2/c^2}\right] \,,
\eeq
where $\mu^2=m^2\kappa^{n-c}/\tau$, with $V=m^2\phi^c$, $T=\tau\phi^n$ 
($c=2$, $n=4$ being of special interest).  As $\mu$ gets large,
the behavior looks more and more like a cosmological constant.
Note that in the light of Eq.~(\ref{eq:tphi}) 
large $\mu$ corresponds to the strong coupling regime
where DBI analysis can be trusted.
The evolution is illustrated in Figure~\ref{fig:phi2}.

Other attractors giving $w\ne-1$ appear in the nonrelativistic limit,
$\gamma=1$.  Here we want $v\to0$, and realizations include the
exponential potential
$V\sim e^{-\lam\kappa\phi}$, 
but with any power law
or less rapid exponential form for $T$.  In particular, we can keep
$T\sim\phi^4$.   In this regime,
\beq
w=-1+\frac{\lambda^2}{3} \,,
\eeq
as for quintessence.  However, if we also take
$T\sim e^{-\lam\kappa\phi}$, 
then $v=\,$constant and we can get a finite value of $\gam$ different
from 1.  The equation of state is
\beq
w=-1+\frac{\lam^2}{6}\,
\frac{\sqrt{\lam^4+12(v-1)\lam^2+36}-\lam^2}{3+(v-1)\lam^2}\,. \label{eq:wexp}
\eeq
We approach the cosmological constant value for $\lam^2\ll1$ or
$v\gg\lam^2$.

We summarize the accelerating attractor solutions in
Table~\ref{tab:attsol}.

\begin{table}[htbp]
\begin{center}
\begin{tabular*}{0.9\columnwidth}
{@{\extracolsep{\fill}} c c c c c c}
\hline
$V/T$ & $\gam$ & $\lam^2/\gam$ & $w$ & $\Omega_{\phi}$ & Stability \\
\hline
$\infty$ & $\infty$ & $0$ & $-1$ & 1 & yes \\
$\infty$ & $\infty$ & const & $-1+\lam^2/(3\gam)$ & 1 & yes \\ 
const & const & const & Eq.~(\ref{eq:wexp}) & 1 & $\lam^2\le3\gam$ \\
0 & 1 & const & $-1+\lam^2/3$ & 1 & $\lam^2\le3$ \\ 
\hline
\end{tabular*}
\end{center}
\caption{Summary of accelerating attractor properties.  The columns
give the values of the quantities for the attractor solution, and
the stability criteria.  For $\lam^2/(3\gam)\ge1$, either the field
switches to the solution with $w=0$ or no attractor exists.
Quintessence attractors can only access the class represented by
the last row.}
\label{tab:attsol}
\end{table}

\section{Comparison with Observations \label{sec:meta}}

While the attractor solutions bring the field to a cosmological
constant behavior or near to it, this could be in the future.  We
need to consider whether DBI theory is consistent with the current
observations.  Without going into great detail, our conclusion is
that generally it is.  The boost factor during the matter dominated
era is driven toward unity, so from Eq.~(\ref{eq:wdef}) we have
$w\to-1$.  Thus we reach an early time ``frozen'' state, looking like
the cosmological constant, for a wide range of initial conditions
including relativistic $\gamma$ (cf.\ Fig.~\ref{fig:gamlp}).

The field then evolves away from the frozen state along the same
generic thawing trajectory as quintessence, $w'=3(1+w)$
\cite{caldwelllinder,cahndl} -- recall
that when $\gamma\approx1$, DBI becomes quintessence-like.  Since
the attractor solution will pull the field back toward $w\approx-1$,
the trajectory often does not deviate far from $w=-1$ for a
range of potential parameters.

Taking as a concrete example the potential as in Fig.~\ref{fig:gamlp}, 
with $\phi_i=1$,
the distance to the cosmic microwave background
last scattering surface, $d_{\rm lss}$, agrees with the standard
cosmological constant cosmology $\Lambda$CDM to 0.67\%, for the same
present matter density.  Other values
of $\phi_i$ give even smaller deviations, and the agreement
improves as $\mu$ increases.
For the potential as in Fig.~\ref{fig:phi2}, with $\phi_i=0.2$,
the agreement is 1.2\%, and again improves as $\phi_i$ or $\mu$ increases.
Considering distances to redshifts $z\le2$, e.g.\ as measured by Type Ia 
supernovae 
(see, for example, \cite{kowalski}), the deviation from $\Lambda$CDM
is at worst 1.7\% and 2.9\% respectively.  (Note the $\mu=20$
case of Fig.~\ref{fig:phi2} gives 0.28\% agreement on $d_{\rm lss}$ and
0.71\% on $z\le2$ distances.)  All these deviations are within
current observational constraints, although future data will be able
to place increasingly tight lower bounds on the effective mass $\mu$.

\section{Conclusions \label{sec:concl}}

We have shown that DBI string theory can achieve
dynamics approaching the cosmological constant and obtaining agreement
with cosmological observations.   These are attractor solutions that
substantially ameliorate the fine tuning of initial conditions.
Several of the accelerating classes cannot be realized within quintessence,
but instead arise from the relativistic nature of the DBI action with
its Lorentz factor $\gamma$.

Unlike quintessence, standard renormalizable potentials like those
with an $m^2\phi^2$ term exhibit attractor behavior.
The linear potential is one example
that has an attractor to a future de Sitter state.
For a range of reasonable masses and coupling values such models are
viable under the current cosmological observations such as
distance-redshift data.

Also unlike quintessence this approach starts from a fundamental basis
in string theory. 
The DBI action arises as the low energy effective theory describing the
dynamics of a probe D3-brane. 
The results, including correspondence to the cosmological
constant, hold with the natural form of the brane
tension $T\sim\phi^4$, but also if it is distorted; 
they  also hold taking into account the breaking of conformal invariance 
and the generation of a mass term in the potential. 
The important property is the ratio $V/T$.  Given this,
the relativistic kinetic properties of the DBI action
allow cosmological constant or $w\approx-1$
states to be realized with some degree of naturalness.

Increasingly accurate cosmological data will be able to test
directly aspects of fundamental string theory within the DBI framework.
Such connections between
string theory and astrophysical data offer exciting prospects for
revealing the nature of the cosmological constant and the accelerating
universe.

\acknowledgments

This work has been supported by the World Class University grant
R32-2008-000-10130-0.  EL has been supported in part by the
Director, Office of Science, Office of High Energy Physics, of the
U.S.\ Department of Energy under Contract No.\ DE-AC02-05CH11231.

\end{document}